\documentclass[showpacs, preprintnumbers,  aps, nofootinbib, prd, 10pt, showkeys, notitlepage, twocolumn]{revtex4-1}

\usepackage{graphicx,amssymb,amsmath,amsthm,amsfonts,epsfig,float,bm,mathtools}

\usepackage[linktocpage]{hyperref}
\usepackage[usenames,dvipsnames]{color}
\usepackage{epstopdf}
\usepackage{aas_macros}
\usepackage{pifont}
\definecolor{darkred}{rgb}{0.5,0,0}
\definecolor{darkgreen}{rgb}{0,0.5,0}
\definecolor{darkblue}{rgb}{0,0,0.5}
\definecolor{prussian}{rgb}{0.0, 0.19, 0.33}
\definecolor{richelectricblue}{rgb}{0.03, 0.57, 0.82}
\definecolor{teal}{rgb}{0.0, 0.5, 0.5}
\definecolor{mediumseagreen}{rgb}{0.24, 0.7, 0.44}
\definecolor{lust}{rgb}{0.9, 0.13, 0.13}
\definecolor{ballblue}{rgb}{0.13, 0.67, 0.8}
\definecolor{darkcyan}{rgb}{0.0, 0.55, 0.55}
\definecolor{mountainmeadow}{rgb}{0.19, 0.73, 0.56}
\definecolor{palecarmine}{rgb}{0.69, 0.25, 0.21}
\definecolor{richcarmine}{rgb}{0.84, 0.0, 0.25}
\definecolor{tangelo}{rgb}{0.98, 0.3, 0.0}
\definecolor{venetian}{rgb}{0.784,0.031,0.082}
\definecolor{bdfrance}{rgb}{0.192,0.549,0.906}

\hypersetup{colorlinks=true, citecolor=venetian,linkcolor=bdfrance, urlcolor=lust}
\usepackage{amsmath,amssymb}
\usepackage{tensor}
\usepackage{mathtools}
\usepackage{amsbsy}
\usepackage{bm}
\usepackage{float}

\newcommand{\be}{\begin{equation}}
\newcommand{\ee}{\end{equation}}
\newcommand{\bear}{\begin{eqnarray}}
\newcommand{\eear}{\end{eqnarray}}

\newcommand{\cL}{{\cal L}}

\newcommand{\cT}{{\cal T}}

\newcommand{\cM}{{\cal M}}

\newcommand{\re}{{\rm e}}

\begin{document}

\begin{center}
{\small The following article has been accepted by American Journal of 
Physics. After it is published, it will be found at: https://publishing.aip.org/resources/librarians/products/journals/}
\end{center}

\title{A dimensional analysis path to $h$ and the Bohr atom structure}

\author{Kostas Glampedakis}
\email{kostas@um.es}

\affiliation{Departamento de F\'isica, Universidad de Murcia, Murcia, E-30100, Spain}


\begin{abstract}

Traditionally, the Planck constant $h$ makes its debut appearance in quantum physics textbooks in the context of the blackbody radiation law and 
subsequently as a fundamental ingredient of the physics of the photoelectric effect and the Bohr atom. In this paper we consider an alternative 
timeline path (which could have taken place several years before the proposal of the Bohr atom) where a classical physics hydrogen atom is studied with dimensional 
analysis techniques in combination with  empirical laws of blackbody radiation. The outcome of this ``classical physicist's'' approach 
is the identification of the correct fundamental Planck constant and the reconstruction of the energy and size scales of the Bohr atom. 

\end{abstract} 
  
\maketitle


\section{Introduction}

It is not unusual for physicists to entertain themselves with  ``alternative timeline" scenarios where a major discovery or breakthrough takes place
in a different way~\cite{rindler95}. Such mental excursions have scientific and educational value on their own right by highlighting paths or plausible 
shortcuts not having been taken or having been overlooked by the real life protagonists. 

The present paper is concerned with a situation of this kind by considering the ``rediscovery'' of the Planck constant $h$ 
and the basic Bohr model structure of the hydrogen atom  (that is, its characteristic energy and size) via the use of standarised 
methods of dimensional analysis~\cite{DAbook1, DAbook2, DAbook3,  DAwebsite, price03}.  

Our analysis adopts the point of view that a physicist might have if they had not seen the development of quantum mechanics but they were
familiar with some of  the most famous open problems of physics at the turn of the previous century, namely, the blackbody's thermal radiation law, 
the photoelectric effect and the atomic structure. In addition our physicist may have been familiar with the latest developments in dimensional analysis, 
namely, the discovery of the so-called Buckingham $\Pi$ theorem~\cite{DAbook1, DAbook2, DAbook3, DAwebsite}  (hereafter termed the ``$\Pi$ theorem") 
which took place around the same period. 

As a working hypothesis we can imagine that the objective of that physicist would have been the dimensional analysis of the hydrogen atom at the level of classical physics. 
Not surprisingly, and as we discuss below, this initial attempt leads to predictions that are in stark disagreement with
experimental atomic physics data. Faced with this failure, our imaginary physicist would have welcomed any input from studies of the other two problems 
(blackbody radiation and photoelectric effect) since these too are manifestations of the properties of matter and its interaction with the electromagnetic field. 
It is the combined dimensional analysis study of all three systems that eventually permits the determination of a new ``Planck constant" and the 
atomic energy and size scales.

Notation and conventions: we use $\{ \cM,\cL,\cT \}$ to denote the three fundamental dimensions of mass, length and time while $[X]$ denotes the dimensionality
of the magnitude $X$ (e.g. for the speed  of light we have $[c] = \cL \cT^{-1}$). Numerical factors are omitted in all dimensional relations since the method is 
intrinsically oblivious  to them. We adopt Gaussian electromagnetic units; with this choice Maxwell's theory contains the single constant $c$ and no new unit/dimension
is required for the charge. 


\section{Atomic physics (and dimensional analysis) circa 1900}
\label{sec:setup}

The state of art atomic model for the period discussed in this paper was the one championed by J.J. Thomson, following his discovery  
of the electron\footnote{The name ``electron'' was coined by Stoney in 1891; Thomson instead used the term ``corpuscle'' for the particle he discovered. 
In this paper we have opted for the former term which we believe it to be more suitable for the modern reader. }
in 1897 (see Refs.~\cite{smith97,kragh97} for reviews). 
This so-called ``plum pudding" atom was supposed to be a charge-neutral system, comprising a number of electrons embedded  in a positive charge matrix. 
The hydrogen atom was expected to have a single electron located at the center.  Atomic emission spectra were believed to originate from the vibrational motion 
of the electrons about their equilibrium positions\footnote{This is H. Lorentz's electron oscillator model and it is a familiar topic in textbooks of  electromagnetism.}

By 1900 Thomson had measured the electron's charge to mass ratio $e/m_\re$ with reasonable accuracy using three different methods. Similar experiments
were performed by other physicists around the same time~\cite{smith97}. In addition, in 1898 and again in 1903 -- several years before Milikan's famous oil drop 
experiment  -- Thomson published the results of measurements of the elementary charge $e$. These were of the same order of magnitude as the modern 
value~\cite{smith97} and allowed the parallel estimation of the electronic mass:
$e/m_\re \approx 2 \times 10^{16}\, \mbox{esu} /\mbox{g}$, $e \approx 6.5 \times 10^{-10}\, \mbox{esu}$.

 Much progress in dimensional analysis took place more or less in the same period as the aforementioned developments in atomic physics. See Ref.~\cite{DAhistory}
 for a detailed historical review. This work followed in the footsteps of earlier 19th century physicists like Poisson, Fourier and Maxwell. In his 1873  textbook 
 ``A Treatise on Electricity and Magnetism" Maxwell considers the dimensions of charge and mass.
 The $\Pi$ theorem itself was first proved by Bertrand in 1878 in the context of electrodynamics and heat conduction.  The theorem became
widely known as a practical tool, under the name ``the method of dimensions", due to the work of Rayleigh (in particular his textbook on the theory of 
sound~\cite{rayleighbook}). The theorem was further generalised in 1892 by the French mathematician Vaschy. It was apparently forgotten and rediscovered, independently, 
in 1911 by Riabouchinsky and  in 1914 by Buckingham who also introduced the $\Pi$ symbol for the dimensionless variables. 
Therefore this paper's use of the name ``$\Pi$ theorem" is a minor anachronism which, nevertheless, we adopt for the benefit of the modern reader.

In a nutshell the $\Pi$ theorem states that given a system described by $n$ parameters with $m$ independent dimensions, we can build $n-m$ independent dimensionless 
$\Pi_i$ ratios (with $ i \in [1,n-m] $) which are physically related. This can be solved for a given $\Pi_j$ so that $ \Pi_j =  f ( \Pi_{i \neq j} )$ for some
undetermined function $f(x)$.

Circa 1900, then, the scene was fully set for the application of this novel dimensional analysis technique to the problem of atomic structure.


\section{The hydrogen atom through the prism of dimensional analysis}

\subsection{The dimensional analysis of the ``classical" hydrogen atom}

In this first part of our analysis, the hydrogen atom is assumed to be an entirely classical system, something along the lines of Thomson's pudding model. 
In its equilibrium state the atom is expected to be an \emph{electrostatic} system comprising two opposite charges $\pm e$ with characteristic energy $E_0$ and 
length scale $a_0$.  The ``moving part" of this atom  is the electron of mass $m_\re$ while the much heavier positive charged component is assumed to be stationary in the 
center of mass frame. 

Without Planck's constant in our toolkit, the available dimensional analysis parameters  -- apart from $E_0$ and $a_0$ -- are those of classical physics i.e.~$\{ m_\re, c, e \}$ 
where $c$ is the speed of light. 
The first two of these have obvious dimensionalities while for the dimensionality of charge we use the electrostatic energy expression $E_0 \sim e^2/r$ to find
$[e] = \cM^{1/2} \cL^{3/2} \cT^{-1}$. 

According to the $\Pi$ theorem, having the four parameters $\{ E_0,  m_\re, c, e \}$ with all three basic dimensions in them means that there is a single independent $\Pi$ ratio. 
This is $ \Pi = E_0 / m_\re c^2$ and therefore the unique dimensional analysis expression for the energy is,
\be
E_0 = m_\re c^2. 
\label{E0in}
\ee
(This might have been identified as the electron's mass-energy but special relativity was unknown for most of the time period discussed in this paper.)
%
A similar analysis for the atomic lengthscale $a_0$ leads again to a unique result, the so-called electron classical radius~\cite{EM1}
\be
a_0 =  \frac{e^2}{m_e c^2}.
 \ee
The same relation follows from $E_0 = e^2 /a_0$. (Indeed, an electromagnetic origin for the electron mass was a popular idea in the 1900s.) 
With the experimentally obtained $m_\re$ and $e$ available around 1900, both parameters could have been numerically evaluated.
 Using the early 1900s experimental results quoted in Sec.~\ref{sec:setup} we find $E_0 \approx 3 \times 10^{-5} \, \mbox{erg} \approx 20\, \mbox{MeV} $
 (approximately $40$ times higher than the modern value) and $a_0 \approx 1.4 \times 10^{-14}\,\mbox{cm} \approx 0.1 \,\mbox{fm}$.

Of course, both results fall well off the mark:  $E_0$ lies far above the experimental energy scale ($\sim$ 10 eV) of atomic physics and chemistry. 
Similarly,  the predicted $a_0$ is far smaller than the expected atomic size ($\sim e^2/E_0 \sim 10^{-8}\,\mbox{cm} $) of the Thomson atom, 
being comparable to the electron classical radius. This is the same length scale that marks the
domain of validity of classical electromagnetism\footnote{The length scale $a_0$ and the associated time scale $a_0/c$ appear in  the equation of motion of a 
point charge in an exterior electromagnetic field that takes into account the self-force on the body due to the emission of radiation. 
This Lorentz-Dirac equation leads to unphysical solutions unless the body is restricted to have a size $\gg a_0$ 
and the fields are restricted to vary on time scales $ \gg a_0/c$. See Refs. \cite{EM1,EM2, EM3} for detailed discussions.}.

The conclusion reached by our imaginary physicist would have been clear: classical physics utterly fails when summoned to describe the scale 
of the hydrogen atom's energy and size. 


\subsection{The ``universal light" of blackbody thermal radiation}
\label{sec:bb}

The blackbody radiation law occupies a special place in the history of physics as it came to be associated with the groundbreaking concept of quantisation. 
Planck's famous radiation formula was the culmination of an effort that had lasted several decades, starting with Kirchhoff's work on the emission and absorption of
radiation. 

In 1896 Wien found an empirical thermal radiation law which was in convincing agreement with observations for all frequencies: 
\be
J (\lambda, T) = c_1 \frac{T}{\lambda^5} e^{-c_2/\lambda T},
\label{Iwien}
\ee
where $J$ is the spectral radiation intensity at wavelength $\lambda$ and temperature $T$, see Ref.~\cite{UVmyth} for a more detailed discussion of this and the 
following formulae. The phenomenological constants $c_1, c_2$ do not depend on the black body's material properties. 

More famous is Lord Rayleigh's theoretical formula for $J$ (published in the spring of 1900) which he derived from classical statistical mechanics
\be
J (\lambda, T) =  c_3  \frac{T}{\lambda^4},
\label{Irayleigh}
\ee
with $c_3 = 2 c k_{\rm B}$. This formula was supposed to apply to the long-wavelength regime; when pushed beyond its domain of validity 
it is known to lead to the ultraviolet catastrophe divergence.  

Further experiments in late 1900 revealed that Wien's law~\eqref{Iwien} was inaccurate in the low frequency regime. Indeed the $\lambda \to \infty$ limit 
of~\eqref{Iwien} does not coincide with~\eqref{Irayleigh}. 

The final chapter to this story was written by Max Planck who  presented his famous formula in October 1900:
\be
J (\lambda, T) = \frac{c_4}{\lambda^5} \left ( e^{c_5/\lambda T} -1 \right )^{-1},
\label{plancklaw}
\ee
with yet another set $c_4, c_5$ of matter-independent constants. This empirical formula, which provided an excellent fit to the experimental data, reduces to~\eqref{Irayleigh}  
when $\lambda \to \infty$.

What  would have been important for our classical physicist is the observation that in all cases the spectral intensity of thermal radiation is described by a function
\be
J (\lambda, T) = f( T, \lambda,   e^{ \Delta/\lambda T} ),
\label{plancklaw2}
\ee
which necessarily comes with the introduction of a new ``universal" constant $\Delta$. It is more convenient to work with the energy $E = k T$ instead of the temperature 
so we can redefine the constant 
\be
\frac{ \Delta}{\lambda T} \to \frac{ \Delta}{\lambda E}.
\ee
The new constant $ \Delta$ has dimensions $[ \Delta] = [\lambda] [E] = \cM \cL^3 \cT^{-2}$; its numerical value can be obtained by fitting~\eqref{plancklaw} to experimental 
blackbody spectra. In this way Planck measured the rescaled constant $\Delta/c$, which  became the eponymous constant $h$ (see Sec.~\ref{sec:Hrevisit} for more details). 
As far as the relation~\eqref{plancklaw2} is concerned though, any  $c$-rescaled $ \Delta$ is as fundamental  as $ \Delta$ itself.


\subsection{The hydrogen atom revisited}
\label{sec:Hrevisit}

Returning to the dimensional analysis of the classical hydrogen atom, we augment our list of parameters with the new blackbody constant $ \Delta$;
an early 1900s physicist might have been compelled to do so believing it to be a constant of truly fundamental nature that has something to do with the interaction 
between matter and radiation.
Then, we can write the following dimensional relations:
\bear
&& E_0 = e^\alpha m^\beta_\re c^\gamma  \Delta^\delta,
\label{Eeq1}
\\
&& a_0 = e^\sigma m^\zeta_\re c^\theta  \Delta^\lambda.
\label{a0eq1}
\eear
Balancing the dimensions of both sides of~\eqref{Eeq1} 
\begin{align}
\cM \cL^2 \cT^{-2}  &=   \left ( \cM^{1/2} \cL^{3/2} \cT^{-1} \right )^\alpha \cM^\beta  \left ( \cL \cT^{-1} \right )^\gamma  
 \nonumber \\
&\quad \left ( \cM \cL^3 \cT^{-2} \right  )^\delta,
\end{align}
leads to the algebraic system
\be
\beta + \frac{1}{2} \alpha + \delta =1, \quad \frac{3}{2} \alpha + \gamma + 3\delta =2, \quad \alpha + \gamma  + 2\delta  = 2.
\ee
This can be solved  for three of the four parameters
\be
\alpha = -2\delta, \quad \beta = 1, \quad \gamma = 2.
\ee
Repeating the same procedure for~\eqref{a0eq1} we find
\be
\zeta + \lambda  + \frac{1}{2}\sigma  = 0, \quad \theta + 3\lambda + \frac{3}{2} \sigma = 1, \quad 
 \theta + 2\lambda +  \sigma = 0.
\ee
(The same system of equations can be obtained if we combine $E_0 = e^2 /a_0$ with the above solutions for $\alpha,\beta, \gamma$.)
This is solved by
\be
\theta = -2, \quad \zeta =-1, \quad \sigma = 2 (1-\lambda).
\ee
Inserting the solutions back into~\eqref{Eeq1}-\eqref{a0eq1}
\be
 E_0 = m_\re c^2 e^{-2\delta}   \Delta^\delta, \qquad   a_0 = \frac{e^{2(1-\lambda)}  \Delta^\lambda }{m_\re c^2}.
 \label{Ea0eq2}
\ee
Dimensional analysis has brought us this far. In order to proceed further some additional assumption is needed. 
For example, the value $\delta=0$ can be discarded because an electrostatic energy should depend on the charge.  
Also, it is reasonable to expect that $c$ should not appear in the final expressions because the system is electrostatic rather than electrodynamic 
(in the latter case we would expect it to be radiatively unstable). This is achieved by  setting $\lambda = -\delta$ and  introducing the rescaled constant
\be
 \Delta = H c^{-2/\delta}.
\ee
The resulting atomic energy and radius are:
\be
 E_0 = m_\re  \left ( \frac{H}{ e^2} \right )^\delta , \qquad a_0 = \frac{e^2}{m_\re } \left ( \frac{e^2}{H} \right )^\delta.
 \label{Ea03}
\ee
Comparing the energy expression with~\eqref{E0in}, it is easy to see that the parameter
\be
x \equiv \frac{e^2 c^{2/\delta}}{H},
\label{xratio}
\ee
must be dimensionless. 
In the end we are left with a single free parameter $\delta$; without it we cannot determine the dimensionality of $H$. 
All $\delta > 0$ values can be eliminated because they would imply a divergent energy for $e \to 0$.  Therefore we expect $\delta$ to be negative.
Furthermore, recalling that in electrostatics the minimum power of charge is $e^2$ and noting that in~\eqref{Ea03} the charge appears as $e^{-2\delta}$ and $e^{2 (1+\delta)}$,  
we might expect that $\delta = n-1$ with $n$ a non-negative integer. (This assumption is justified more rigorously below, where we apply the full force of the $\Pi$ theorem.)

For $\delta=-1$ the atomic radius becomes independent of $e$. On these grounds we may reject this value and move on to the next choice 
$\delta =-2$. This leads to
\be
E_0 = \frac{m_\re e^4}{H^2}, \qquad  a_0 = \frac{H^2}{m_\re e^2 }.
\label{E0a0final}
\ee
We have no reason to reject these formulae so we put them forward as the final dimensional analysis prediction of our classical physicist. 
In addition the value $\delta =-2$ fixes $[H]$; according to~\eqref{xratio} we have $[H] = [e^2] [c]^{-1}  = [E_0]  \cL  [c]^{-1} = [E_0] \cT$ so that dimensionally $[H] = [h]$. 
That is, it is the constant  $H = h$ and not some other $c$-multiple of $ \Delta$ that emerges as the truly fundamental one in the context of the hydrogen atom. 
With $[H]$ known, we can finally identify the expressions~\eqref{E0a0final} as the characteristic energy and size of a classical hydrogen atom.

As pointed out earlier, the numerical value of $H$ can be obtained with good precision from blackbody data.
(In his seminal 1901 paper~\cite{planck01} Max Planck quotes the numerical value $ h = 6.55 \times 10^{-27}\, \mbox{erg}\, \mbox{sec}$.)
This also allows the numerical estimation of the ``fine structure constant" $x = e^2 /c h \approx 0.01$.

The preceding basic dimensional analysis calculation can be refined via the application of the $\Pi$ theorem.  
For the problem at hand we consider the dimensionless ratios:
\be
\Pi_1 = \frac{E_0}{m_\re c^2}, \quad \Pi_2 = x,
\ee
and the resulting dimensional solution is
\be
\Pi_1 = f (\Pi_2) ~\Rightarrow ~ E_0 = m_\re c^2 f (x),
\label{E0eq4}
\ee 
where the smooth function $f(x)$ is unspecified. From $E_0 = e^2 /a_0$ we obtain a second dimensional expression
\be
a_0 = \frac{e^2}{m_\re c^2 } \frac{1}{f(x)}.
\label{a0eq4}
\ee
Our previous results~\eqref{Ea03} can now be seen as leading-order expressions of these
relations after a Taylor series expansion of $f(x)$ for $x \ll1$. The leading-order power in this series must be $ n = -\delta$ so that the $c$-dependence drops out. 
That is, $f(x) = x^2 [ 1 + {\cal O} (x) ]$ and 
\begin{align}
 E_0 &= m_\re    \left ( \frac{e^2}{H}  \right )^{-\delta} \left [ \, 1 + {\cal O}  (x)  \, \right  ],
 \\
  a_0 &= \frac{e^2}{m_\re}  \left ( \frac{e^2}{H}  \right )^{-\delta}  \left [ \, 1 + {\cal O} (x)   \, \right ]. 
\end{align}
This approach explains why we were allowed to consider only integer values of $\delta$ in~\eqref{Ea03}; in addition, it provides information on the next-order
correction terms (which necessarily are of relativistic nature as they depend on $c$). 


\subsection{Another strand: the  photoelectric effect}

The experimental study of the photoelectric effect can be summed up in the following linear fit formula: 
\be
K_\re (f) = A f - \Phi_\re,
\ee
where $K_\re$ is the extracted electron's kinetic energy, $\Phi_\re$ is the material's ionisation threshold, $f$ is the electromagnetic field frequency and $A$ is a 
material-independent (hence ``universal") constant. Our imaginary classical physicist would have been tempted to relate the new dimensionless ratio $A f /E$
to the similar ratio $ \Delta / \lambda E$ of the blackbody radiation law and derive the dimensional relation
\be
A  =  \Delta/c =  H  c^{-1-2/\delta}.
\ee
The two constants $A$ and $H$ are identical for $\delta = -2$, the same value found via dimensional analysis. In fact, we could have arrived at the earlier 
results~\eqref{Ea03} by working with $A$ instead of $ \Delta$. Both options would have been available to an early 1900s physicist.

\section{Conclusions}

We have discussed how the dimensional analysis of a classical hydrogen atom model, in combination with the phenomenological constant of the empirical 
blackbody radiation law (or of the photoelectric effect formula) provides a path to the `discovery' of the Planck constant and the appropriate scales for the
atomic energy and size. All the necessary theoretical and experimental input for such analysis was in place roughly a decade before the actual Bohr model
was proposed (although the key quantisation property of that model could not have been predicted by dimensional analysis). 
In reality things did not play out according to the `timeline' discussed in this paper. By the time  Bohr put forward his atomic model in 1913,
both the Planck constant and the light quanta were more or less familiar concepts from the study of blackbody physics
and the photoelectric effect, while Rutherford's scattering experiments had favoured a ``solar-system" atomic model instead of the Thomson model. 
It is interesting to note though that another dimensional analysis path towards the discovery of $h$ might have been possible a few years before the timeline
considered so far, based solely on the physics of blackbody radiation and without any reference to atomic physics. This is presented in the Appendix.


\appendix

\section{Discovering $h$ in 1893}
\label{sec:appendix}

Here we show that, with a slightly anachronistic use of dimensional analysis for the blackbody's total radiated intensity $I$, a classical physicist could have discovered the 
Planck constant as early as 1893, well before the appearance of the various blackbody spectral intensity formulae presented in Section~\ref{sec:bb}. 

In 1893 Wilhelm Wien discovered his eponymous law relating the blackbody temperature to the wavelength $\lambda_{\rm max}$ of maximum
spectral radiation intensity:
\be
\lambda_{\rm max} T = b, 
\label{wienlaw}
\ee
where $b$ is an empirical constant independent of the  blackbody's material. Wien's derivation was based on classical thermodynamics; the same is true for the 1884
derivation of the Stefan-Boltzmann law:
\be
I = \sigma T^4,
\label{SB}
\ee 
with $\sigma$ another empirical ``universal" constant.

The initial dimensional analysis of~\eqref{SB} is based on  just three parameters, namely, $ \{ I, c, k T \} $. (The obligatory thermodynamical pairing of $k$ and $T$ is dictated by 
the energy-temperature  defining relation $T = E/k$; the absence of material-related parameters is a reflection of the universal character of blackbody radiation.)
These parameters, however, cannot even be combined dimensionally, in accordance  with the $\Pi$ theorem (i.e. there are $3-3 = 0$ dimensionless ratios). 
Therefore, the \emph{necessity} of the introduction of  a new constant -- which should be of fundamental character as far as blackbody physics is concerned -- is imposed 
by basic dimensional reasons. 

Let us call the new constant $\Sigma$; the new list of parameters $  \{ I, c, k T,  \Sigma \} $ leads to the unique dimensionless ratio
\be
\Pi  = \frac{I}{ c^\beta (k T)^\gamma \Sigma} ~\Rightarrow ~ I = c^\beta \Sigma (k T)^\gamma,
\ee
where have assumed linearity in $ \Sigma$ without any loss of generality (that is, we have rescaled 
$\Sigma^\alpha \to  \Sigma$ for some power $\alpha$). 
By comparing  to the Stefan-Boltzmann law~\eqref{SB} we can fix $\gamma=4$. The updated dimensional relation can be used to obtain $[\Sigma]$ (recall that $I$
represents energy per unit time and unit surface area):
\be
 I = c^\beta  \Sigma (k T)^4 ~ \Rightarrow ~ [ \Sigma ]  = \cL^{-8-\beta} \cT^{5+\beta} \cM^{-3}.
 \label{Hdim1}
 \ee
The second input in this analysis is provided by Wien's law~\eqref{wienlaw} which we write in the slightly different form
\be
\lambda_{\rm max} k T = B  \sim  c^\delta \Sigma^\theta.
\label{wien2}
\ee 
This expresses the fact that $B =  k b $ can only depend on the two available constants $c, \Sigma$. From~\eqref{wien2} we get
\be
 [ \Sigma ]^\theta = \cM \cL^{3-\delta} \cT^{\delta -2}.
 \label{Hdim2}
 \ee
The combination of~\eqref{Hdim1} and~\eqref{Hdim2} leads to the following algebraic system of equations
\be
1 =  -3\theta, \quad  3-\delta =  - \theta( 8 + \beta), \quad \delta - 2 = \theta (5+\beta).
\ee
Its solution is $\theta =-1/3$ and $\beta = 1 -3\delta$; the resulting dimensionality of 
$\Sigma$ is $[ \Sigma]=  \cL^{3\delta-9}  \cT^{6-3\delta} \cM^{-3} = [c]^{3\delta} \cM^{-3}  \cL^{-9}  \cT^{6}$ which means that we can write
\be
B \sim  \left (   c^{-3\delta} \Sigma \right )^{-1/3},  \qquad
I \sim c \left ( c^{-3\delta}  \Sigma \right ) (k  T)^4.
\ee
The remaining unknown power $\delta$ can be effectively eliminated by a redefinition $ \Sigma \to \tilde{H} = c^{-3\delta}  \Sigma$. 
In terms of this new constant
\be
 B \sim \tilde{H}^{-1/3}, \qquad  I \sim c \tilde{H}   (k T)^4.
\ee
In fact, the dimensionality $ [\tilde{H}] = \cM^{-3}  \cL^{-9}  \cT^{6}$ implies the presence of  a yet simpler constant:
\be
[\tilde{H}] = \left ( \cM  \cL^{3}  \cT^{-2} \right )^{-3} ~\Rightarrow ~ \tilde{H} \sim ( c h)^{-3}.
\ee
where $h$ has dimensions of action. Our final dimensional analysis results are:
\be
B \sim c h, \qquad  I \sim \frac{1}{c^2 h^3} (k T)^4 \sim \frac{c}{B^3} (k T)^4.
 \ee
 As it was the case with the constant $\Delta$ (see Section~\ref{sec:bb}), in this analysis $B$ and $h$ are equally ``fundamental" constants.
 As far as the scales are concerned, the dimensional analysis results are in full agreement with the $I$ and $\lambda_{\rm max}$ obtained from 
 the Planck blackbody law
 \be
J (\lambda, T) = \frac{2\pi h c^2}{\lambda^5} \left ( e^{h c/\lambda k T} -1 \right )^{-1}. 
\label{plancklaw}
\ee
 

\acknowledgements
The author is grateful to Stefanos Trachanas (Univ. of Crete) for bringing to his attention the calculation discussed in  Appendix~\ref{sec:appendix}.
He is also grateful to the anonymous referees for their comments and suggestions that improved the quality of this paper.

%
%
%


\pagebreak


\end{document}